\documentclass{cidr-2019}

\usepackage{amsmath}
\usepackage{amssymb}
\usepackage{wasysym}
\usepackage{graphicx}
\usepackage{xspace}
\usepackage{algorithm}
\usepackage{algorithmicx}
\usepackage{algpseudocode}
\usepackage{color}
\usepackage{units}
\usepackage{mathrsfs}
\usepackage{cite}

\usepackage{subcaption}

\usepackage{enumitem}
\setitemize{noitemsep,topsep=0pt,parsep=0pt,partopsep=0pt}
\setenumerate{noitemsep,topsep=0pt,parsep=0pt,partopsep=0pt,leftmargin=5mm}

\newcommand{\sparagraph}[1]{\noindent {\bf #1}}

\newcommand{\join}{\bowtie}
\newcommand{\DRL}{DRL\xspace}

\newcommand{\actions}{\mathscr{A}}

\newcommand{\CM}{ \mathscr{M} }

\newcommand{\nextSec}[0]{}

\newcommand{\cut}[1] {}

\newcommand{\totsize}[1] {0.9\textwidth}

\begin{document}
\title{Towards a Hands-Free Query Optimizer through Deep Learning}

\numberofauthors{2}
\author{
  \alignauthor
  Ryan Marcus\\
  \affaddr{Brandeis University}
  \email{ryan@cs.brandeis.edu}
% 2nd. author
  \alignauthor
  Olga Papaemmanouil\\
  \affaddr{Brandeis University}\\
  \email{olga@cs.brandeis.edu}
}

\maketitle

\begin{abstract}
Query optimization remains one of the most important and well-studied problems in database systems. However, traditional query optimizers are complex heuristically-driven systems, requiring large amounts of time to tune for a particular database and requiring even more time to develop and maintain in the first place. In this vision paper, we argue that a new type of query optimizer, based on deep reinforcement learning, can drastically improve on the state-of-the-art. We identify potential complications for future research that integrates deep learning with query optimization, and describe three novel deep learning based approaches that can lead the way to end-to-end learning-based query optimizers. 
\end{abstract}

\section{Introduction}

% readd: opt_rules
Query optimization, e.g. transforming SQL queries into physical execution plans with good performance, is a critical and well-studied problem in database systems (e.g.~\cite{volcano, systemr, robust_qo, quickpick}). Despite their long research history, the majority of existing query optimization systems share two problematic properties:
\begin{enumerate}
\item{They are, or are composed of, carefully tuned and complex \emph{heuristics} designed using many years of developer-based experience. Furthermore, these heuristics  often require even more tuning by expert DBAs to improve query performance on each individual database (e.g.~tweaking optimization time cutoffs, adding query hints, updating statistics, tuning optimizer ``knobs'').}
\item{They take a ``\emph{fire and forget}'' approach in which the observed performance of a execution plan is never leveraged by the optimization process in the future, hence preventing query optimizers from systematically ``learning from their mistakes.''}
\end{enumerate}

Of course, there are several notable exceptions. Many optimizers use feedback from query execution to update cardinality estimates~\cite{AboulnagaSelftuningHistogramsBuilding1999, ChenAdaptiveSelectivityEstimation1994, leo}, and many adaptive query processing systems~\cite{adaptive_qp_rl, cuttlefish} incorporate feedback as well. However, in this vision paper, we argue that recent advances in \emph{deep reinforcement learning} (DRL)~\cite{deep_rl} can be applied to query optimization, resulting in a ``hands-free'' optimizer that (1) can tune itself for a particular database automatically without requiring intervention from expert DBAs, and (2) tightly incorporates feedback from past query optimizations and executions in order to improve the performance of query execution plans generated in the future.

%Reinforcement learning problems generally involve an agent ewhich takes various actions within an environment.  Deep reinforcement learning~\cite{deep_rl} uses 

%Towards realizing the above vision lie a number of research challenges. 
Deep reinforcement learning is a process in which a machine learns a task through continuous feedback with the help of a neural network~\cite{dnn}. It is a iterative learning process where the machine (an \emph{agent}) repeatedly selects actions and receives feedback about the quality of the actions selected. \DRL algorithms train a neural network model over multiple rounds (\emph{episodes}), aiming to maximize the performance of their selected actions (\emph{policies}). This performance feedback, the indicator of whether or not an agent is performing well or poorly, is referred to as the \emph{reward signal}.

While deep learning has been previously applied to database systems (e.g. indexes~\cite{ml_index}, physical design~\cite{selfdrivingcidr}, and entity matching~\cite{deep_entity}), deep \emph{reinforcement} learning has not received much attention. Despite applications in multiple domains~\cite{deep_rl}, applying \DRL algorithms to  query optimization generates a number of research challenges.  First, \DRL algorithms initially perform very poorly, and require extensive training data before achieving competitive performance. Second, it is generally assumed that that the reward signal is cheap to calculate. In query optimization, the most natural performance indicator to use is the query latency. However, training on (and hence executing) large numbers of query plans (especially poorly optimized query plans) and  collecting their latency for feedback as a reward signal to a \DRL agent can be extremely expensive. Using the optimizer's cost model as a performance indicator is also problematic, as cost models are themselves complex, brittle, and often rely on inaccurate statistics and oversimplified assumptions. %Hence, optimizers output a cost metric that is not strongly correlated with actual query execution times. 

Second, the enormous size of the query plan search space for any given query  causes naive applications of \DRL to fail.  For instance, while \DRL  can be used to learn policies that tackle join order enumeration~\cite{rejoin}, training these models to additionally capture physical operator and access path selection dramatically lengthens the training process and hinders  convergence to an effective policy. 

In this vision paper, we describe and analyze  potential solutions to the above challenges, each representing directions for further research that tightly integrates deep learning-based theory with query optimization. We propose two novel \DRL approaches: \emph{learning from demonstration} and \emph{cost model bootstrapping}. The first approach involves initially training a model to imitate the behavior of a state-of-the-art query optimizer, and then fine-tuning that model for increased performance. The second approach involves using existing cost models as guides to help \DRL  models learn more quickly. Finally, we propose and analyze the design space of an \emph{incremental training} approach that involves learning the complexities of query optimization in a step-by-step fashion. %, mirroring the way humans learn algebra before calculus.  

%To support our vision, we begin with a case study by examining ReJOIN~\cite{rejoin}, a proof-of-concept join ordering enumerator powered by deep reinforcement learning. Driven by our experience with ReJOIN, we identify two key challenges for future research towards a fully end-to-end  learning-based query optimizer. 

{We start in Section~\ref{sec:drl} with an brief introduction to \DRL and an overview of a case study \DRL-based join enumerator in Section~\ref{sec:rejoin}. In Section~\ref{sec:challenges}, we detail the three main challenges that \DRL-based query optimizers need to overcome. In Section~\ref{sec:solutions}, we analyze our  proposed future research directions, and we conclude in Section~\ref{sec:conclusions}. }

%We discuss ReJOIN's strengths and weaknesses, and argue that ReJOIN concretely demonstrates the potential for deep reinforcement learning to have a significant impact on query optimizers and on query optimization research.

%Despite ReJOIN's promising results, we identify two key challenges for future research towards a fully end-to-end deep reinforcement learning query optimizer.  First, the enormous size of the ``search space'' (e.g., the number of potential join orderings, physical operators, access paths, etc.) causes naive applications of deep reinforcement learning to fail (e.g., they take too long to converge to a performant policy, or they never converge to such a policy). Second is the \change{optimization metric, or, in reinforcement learning terms, the ``reward signal''.} Deep reinforcement learning algorithms (a) often require thousands of episodes worth of training data before achieving good performance, and (b) initially perform very poorly. A general assumption in deep reinforcement learning research is that the reward signal -- the indicator of whether or not an agent is performing well or poorly -- is cheap to calculate (e.g., a score in a video game). However, while query latency is the most natural reward signal to use for a deep reinforcement learning-based query optimizer, evaluating query plans (especially poorly optimized query plans) can be extremely expensive. 

%%% Local Variables:
%%% mode: latex
%%% TeX-master: "dlopt"
%%% End:
\nextSec
\section{Deep Reinforcement Learning} \label{sec:drl}
%Here, we give a brief overview of \DRL and discuss the lessons we learned from applying \DRL to join ordering enumeration~\cite{rejoin}. 
%Reinforcement learning problems are ones in which an agent exists in a state, and selects from a
%number of actions. Based on the state and action selected, the agent receives a reward and is placed into a new state.
%The agent's goal is to use information about its current state and its past experience to maximize reward over time.

Reinforcement Learning (RL)  \cite{q}  is a machine learning technique that enables an agent to learn in an interactive environment by trial and error using feedback from its own actions and experiences.  More formally, an \emph{agent} interacts with an \emph{environment}. The environment tells the agent its current state, $s_t$, and a set of potential actions $\actions_t = \{a_0, a_1, \dots, a_n\}$ that the agent may perform. The agent selects an action $a \in \actions_t$, and the environment gives the agent a \emph{reward} $r_t$ based on that action. The environment additionally gives the agent  a new state $s_{t+1}$ and a new action set $\actions_{t+1}$. This process repeats until the agent reaches a \emph{terminal state}, where no more actions are available. This marks the end of an \emph{episode}, after which a new episode begins. The agent's goal is to maximize its reward over episodes by learning from its experience (previous actions, states, and rewards). This is achieved by balancing the \emph{exploration} of new never-before-tried actions with the \emph{exploitation} of  knowledge collected from past actions. 

%One of the key challenges in applying RL to a particular domain is ``massaging'' the problem into these terms (i.e., designing its actions, states, and rewards). Once formalized in this way, reinforcement learning techniques aim to improve upon an initially random \emph{policy}, a function that selects an action in a given state.

\sparagraph{Policy Gradient} One subset of reinforcement learning techniques is policy gradient methods~\cite{reinforce}.  Here the agents select actions based on a parameterized \emph{policy} $\pi_\theta$, where $\theta$ is a vector that represents the policy parameters. Given a state $s_t$ and an action set $\actions_t$, the policy $\pi_\theta$ outputs one of the potential actions from $\actions_t$. %Actions are then selected using various methods~\cite{deep_rl}. 
 
 Reinforcement learning aims to optimize the policy $\pi_\theta$ over episodes, i.e., to identify the policy parameters $\theta$ that optimizes the expected reward.  The expected reward that a policy will receive per episode is denoted $J_\pi(\theta)$. A reinforcement learning agent thus seeks the vector $\theta$ that maximizes the reward $J_\pi(\theta)$, but the reward $J_\pi(\theta)$ is typically not feasible to precisely compute. Hence, policy gradient methods search for such a vector $\theta$ by constructing an estimator $E$ of the \emph{gradient} of the expected reward: $E(\theta) \approx \nabla_\theta J_\pi(\theta)$.

Real-world applications require that any change to the policy parameterization has to be smooth, as drastic changes can (1) be hazardous for the system and (2) cause the policy to fluctuate too severely, without ever converging. %any useful initializations of the policy based on domain knowledge would  vanish after a single step. 
For these reasons, given an estimate $E$, \emph{gradient ascent/descent} methods~\cite{sgd}  tune the initial parameters $\theta$ by increasing each parameter in $\theta_i$ by a \emph{small} value when the gradient $\nabla_{\theta_i} J_\pi(\theta)$ is positive (the positive gradient indicates that a larger value of $\theta_i$ will increase the reward), and decreasing the parameters in $\theta_i$ by a small value when the gradient is negative.

\sparagraph{Deep Reinforcement Learning} In \DRL,  \emph{policy gradient deep learning methods} (e.g., ~\cite{ppo, trpo}) represent the policy $\pi_\theta$ as a neural network, where $\theta$ is the network weights. The policy is improved by adjusting the weights of the network based on the reward signal from the environment. Here, the neural network  receives as input a representation of the current state, and transforms it through a number of hidden layers.  Each  layer transforms (through an activation function) its input data and and passes its output  to the subsequent  layer. Eventually, data is passed to the final action  layer. Each neuron in the action layer represents an action, and these outputs are normalized to form a probability distribution. The policy  selects actions by sampling from this probability distribution, aiming to  balance exploration and exploitation. Selecting the \emph{mode} of the distribution instead of sampling from the distribution would represent a \emph{pure exploitation} strategy. Choosing an action uniformally at \emph{random} would represent a \emph{pure exploration} strategy.

%%% Local Variables:
%%% mode: latex
%%% TeX-master: "dlopt"
%%% End:
  
\nextSec

\section{Case Study: ReJOIN}
\label{sec:rejoin}

One of the key challenges in applying RL to a particular domain is ``massaging'' the problem into the terms of reinforcement learning (i.e., designing its actions, states, and rewards). In this section, we present a case study of ReJOIN, a deep reinforcement learning join order enumerator. We first give a brief overview\footnote{\small Details about ReJOIN can be found in~\cite{rejoin}.} of ReJOIN, and highlight key experimental results.  While ReJOIN focused exclusively on join order enumeration (it did not perform operator or index selection), it represents an example of how query optimization may be framed in the terms of reinforcement learning. 

\sparagraph{Overview} ReJOIN performs join ordering in a bottom-up fashion, modeling the problem in the terms of reinforcement learning.  Each query sent to the optimizer represents an episode, and ReJOIN learns over multiple episodes (i.e., continuously learning as queries are sent). Each state represents subtrees of a binary join tree, in addition to information about query join and selection predicates. Each action represents combining two subtrees together into a single tree. A subtree can represent either an input relation or a join between subtrees. The episode ends when all input relations are joined (a terminal state). At this point, ReJOIN assigns a reward to the final join ordering based on the optimizer's cost model. The final join ordering is sent to the optimizer to perform operator selection, index selection, etc., and the final physical plan is executed.

Intuitively, ReJOIN uses a neural network to iteratively build up a join order. When the optimizer's cost model determines that the resulting query plan (using the join ordering selected by ReJOIN) is good (i.e., a low cost), ReJOIN adjusts its neural network to produce similar join orderings. When the optimizer's cost model determines the resulting plan is bad (i.e., a high cost), ReJOIN adjusts its neural network to produce different join orderings.

\begin{figure}
\centering
\includegraphics[width=0.28\textwidth]{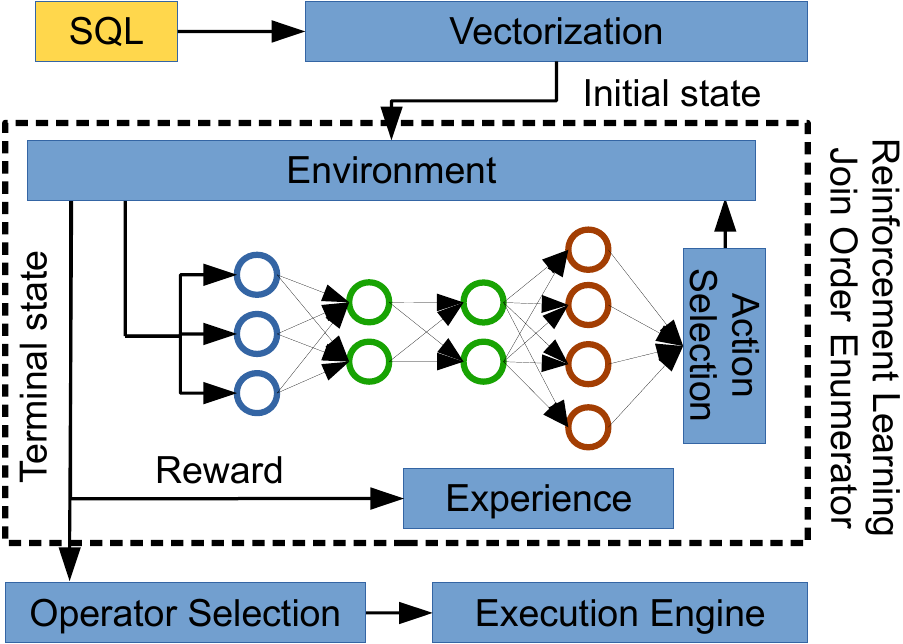}
\caption{\small The ReJOIN Framework}
\label{fig:rejoin}
\end{figure}

\sparagraph{State and Actions} The framework is shown in Figure~\ref{fig:rejoin}. Formally, given a query $q$ accessing relations $r_1, r_2, \dots, r_n$, we define the initial state of the episode for $q$ as $s_1 = \{r_1, r_2, \dots, r_n\}$. This state is expressed as a \emph{state vector}. This state vector is fed through a neural network, which produces a probability distribution over potential actions. The action set $\actions_i$ for any state is every unique ordered pair of integers from $1$ to $|s_i|$, inclusive: $\actions_i = \left[1, |s_i| \right] \times \left[1, |s_i| \right]$. The action $(x, y) \in \actions_i$ represents joining the $x$th and $y$th elements of $s_i$ together. The output of the neural network is used to select an action (i.e., a new join), which is sent back to the environment, which transitions to a new state.  The state $s_{i+1}$ after selecting the action $(x, y)$ is $s_{i+1} = \left(s_i - \{s_i[x], s_i[y]\}\right) \cup \{s_i[x] \join s_i[y]\}$. The new state is fed into the neural network. The reward for every non-terminal state (a partial ordering) is zero, and the reward for an action arriving at a terminal state $s_f$ (a complete ordering) is the reciprocal of the cost of the join tree $t$, $\CM(t)$, represented by $s_f$, $\frac{1}{\CM(t)}$. Periodically, the agent uses its experience to tweak the weights of the neural network, aiming to earn larger rewards.  

\begin{figure}
\centering
\includegraphics[width=0.42\textwidth]{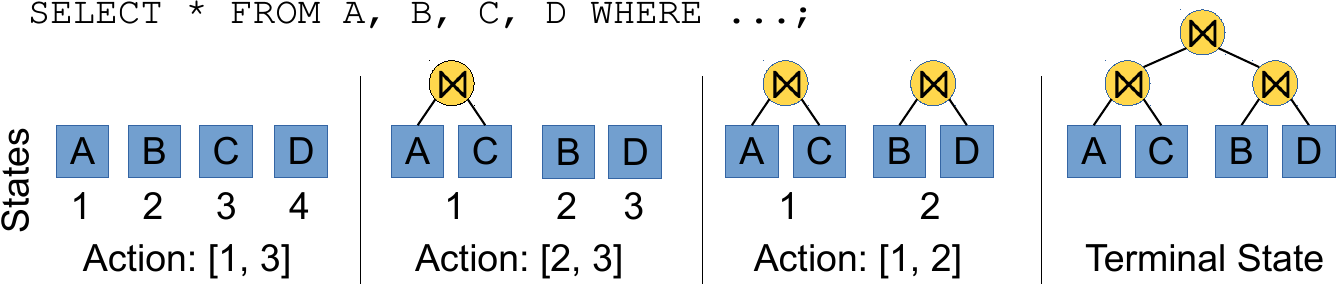}
\caption{\small ReJOIN example}
\label{fig:example}
\vspace{-5mm}
\end{figure}

\sparagraph{Example} Figure~\ref{fig:example} shows an example of this process. Each of the relations in the SQL query are initially treated as subtrees. At each step, the set of possible actions contains every possible pair of subtrees. For example, in Figure~\ref{fig:example}, ReJOIN selects the action \texttt{[1,3]}, so relations $A$ and $C$ are joined. The reward for this action is determined by a DBMS' optimizer cost model. At the next step, ReJOIN selects the action \texttt{[2, 3]}, so relations $B$ and $D$ are joined. Finally, the action \texttt{[1, 2]} is selected, and the $A \join C$ and $B \join D$ subtrees are joined. The resulting state of the system is a terminal state, as no more actions can be selected. The resulting join ordering is sent to a traditional query optimizer, and the optimizer's cost model is used to determine the quality of the join ordering (the reward).

\begin{figure*}
\centering
\begin{subfigure}{0.32\textwidth}
	\centering
	\includegraphics[width=\textwidth]{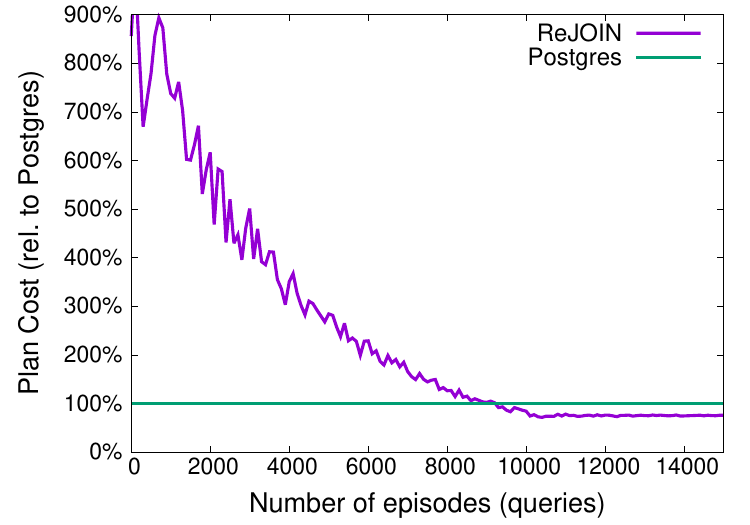}
	\caption{\small ReJOIN convergence}
	\label{fig:convergence}
\end{subfigure}
\begin{subfigure}{0.32\textwidth}
	\includegraphics[width=\textwidth]{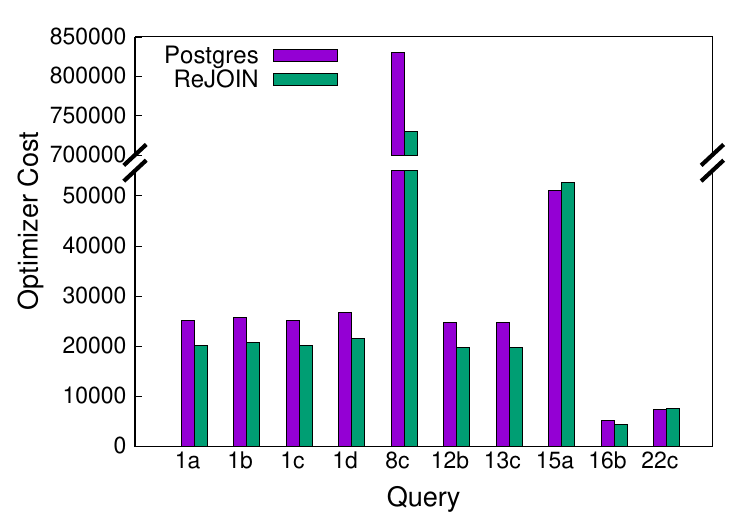}
	\caption{\small Cost of generated plans}
	\label{fig:latency}
\end{subfigure}
\begin{subfigure}{0.32\textwidth}
	\includegraphics[width=\textwidth]{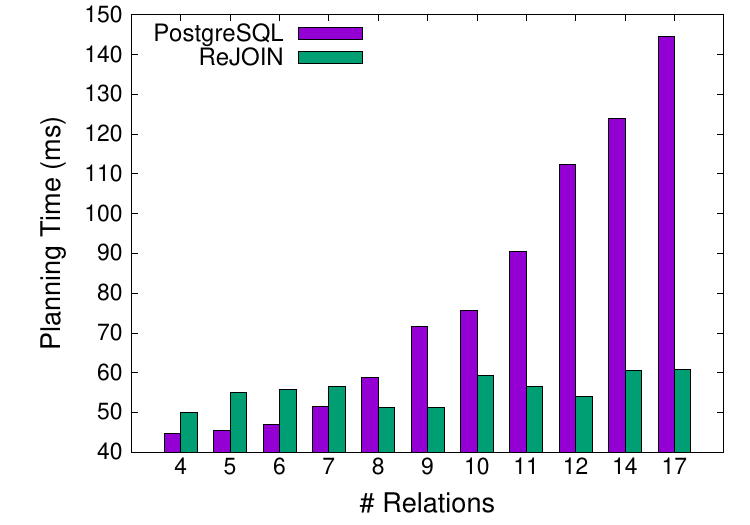}
	\caption{\small Optimization time}	
	\label{fig:plan}
\end{subfigure}
%\vspace{-3mm}
\caption{\small Effectiveness and efficiency results}
\vspace{-3mm}
\label{fig:rejoin_results}
\end{figure*}

\sparagraph{Experimental Results} Figure~\ref{fig:rejoin_results} shows several key experimental results from ReJOIN.
%\begin{enumerate}
{Figure~\ref{fig:convergence} shows the average performance of ReJOIN compared to PostgreSQL during training.  The graph demonstrates that ReJOIN has the ability to learn join orderings that lead to query executions plan with latency close and even better than the ones of PostgreSQL.  However, converging to a good model takes time. Even for the ``limited'' search space of join order enumeration, ReJOIN had to process nearly 9000 queries to become competitive with PostgreSQL.}

{Figure~\ref{fig:latency} shows that the final join orderings selected by ReJOIN (after training) are superior to PostgreSQL according to the optimizer's cost model. While the produced query plans were faster in terms of latency as well~\cite{rejoin}, potential errors in the cost model, and the high human cost of developing and maintaining the cost model, makes directly optimizing for latency much more desirable.}
{Figure~\ref{fig:plan} shows the time required for PostgreSQL and  ReJOIN to select a join ordering. Counter-intuitively, ReJOIN's deep reinforcement learning algorithm (after training) is faster than PostgreSQL's built-in join order enumerator in many cases.}
%\end{enumerate}

\sparagraph{Summary} Our  experiental analysis of ReJOIN~\cite{rejoin} yielded interesting conclusions:
\begin{enumerate}
\item{While ReJOIN is eventually able to learn a join ordering policy that outperforms PostgreSQL (both in terms of optimizer cost and query latency), doing so requires a substantial, but not prohibitive, training overhead}.
\item{ReJOIN's use of a traditional query optimizer's cost model as a reward signal allowed for join orderings to be evaluated quickly. However, this implies that ReJOIN's performance depends on the existence of a well-tuned cost model.}
\item{Counter-intuitively, ReJOIN's \DRL algorithm is faster than PostgreSQL's built-in join order enumerator in many cases. Notably, the bottom-up nature of ReJOIN's algorithm is $O(n)$, where PostgreSQL's greedy bottom-up algorithm is $O(n^2)$.}
\end{enumerate}

ReJOIN is, to the best of our knowledge, the first direct application of deep reinforcement learning to query optimization. Another promising work~\cite{qo_state_rep} has examined how deep reinforcement learning can produce embedded representations of substeps of the query optimization process which correlate strongly with cardinality, with an eye towards a more principled deep reinforcement learning powered query optimizer. Even more recent work~\cite{sanjay_wat} demonstrates how a deep Q-learning~\cite{dqn} approach, with a small amount of pre-training, can perform well when true cardinalities are used as inputs and the optimization target is one of several analytic cost models.

%%% Local Variables:
%%% mode: latex
%%% TeX-master: "dlopt"
%%% End:
\nextSec
\section{Learning-based Query Optimization: Research Challenges} 
\label{sec:challenges}
Inspired by our experience with ReJOIN~\cite{rejoin} as well as other existing work in the area~\cite{qo_state_rep}, we argue that  applications of \DRL theory to query optimization is both promising and possible. However, we next identify three key research challenges that must be overcome in order to achieve an end-to-end \DRL-powered query optimizer. %Afterwards, we propose a number of new research directions to potentially address these challenges.

\sparagraph{Search Space Size}
{While previous work~\cite{rejoin} has demonstrated that reinforcement learning techniques can find good policies in limited search spaces (e.g., join order enumeration in isolation)}, 
%While ReJOIN demonstrates that ability for \DRL algorithms to find good policies in extremely large search spaces (e.g., there are over 500 billion possible join orderings for a query with 10 relations),
  the entire search space for execution plans is significantly larger. {The ReJOIN prototype required 9000 iterations to become competitive with the PostgreSQL optimizer, and in that case only join ordering was considered (no index or operator selection, etc.).} Accounting for operator selection, access path selection, etc. creates such a large search space that approaches from earlier work cannot be easily scaled up. {In fact, a naive extension of ReJOIN  to cover the entire execution plan search space yielded a model that did not out-perform random choice even with 72 hours of training time.} Theoretical results~\cite{near_opt_rl} support this observation, suggesting that adding additional non-trivial dimensions to the problem increases convergence time drastically. %If query latency were used as a reward signal for a fully end-to-end \DRL-driven query optimizer in the style of ReJOIN, convergence time would undoubtedly be prohibitive.

\sparagraph{Performance Indicator}\label{sec:reward_signal}
Deep reinforcment learning algorithms generally make several assumptions about the metric to optimize, i.e., the \emph{reward signal}, that are difficult to guarantee in the context of query optimization. Abstractly, the metric to optimize in query optimization is the latency of the resulting execution plan. However, we next discuss why using  latency as a reward signal leads to two unfortunate complications, namely that the query latency offers neither a \emph{dense} nor a \emph{linear} reward signal.

Many deep reinforcement learning algorithms~\cite{dqn,ppo} assume that, or perform substantially better when, the reward signal is \emph{dense}: provided progressively as the environment is navigated, e.g. each action taken by a reinforcement learning agent achieves some reward.  Furthermore, DRL algorithms often assume that rewards are \emph{linear}, i.e. the algorithms attempt to maximize the sum of many small rewards within an episode. Neither of these assumptions hold in the context of query optimization: query latency is not dense (it can only be measured after a plan has been executed), and it is not linear (e.g., subtrees may be executed in parallel).

%Such a reward function can be difficult to engineer in the context of query optimization, as evaluating the quality of a partial execution plan is a difficult task

%\sparagraph{Dense linear rewards} \change{Abstractly, the metric to optimize in query optimization is the latency of resulting execution plan. However, directly using this latency as a reward signal leads to complications.} Many \DRL algorithms assume that a reward signal is \emph{dense}: provided progressively as the environment is navigated, e.g. each action achieves some reward. Such a reward function can be difficult to engineer in the context of query optimization, as evaluating the quality of a partial execution plan is a difficult task. 
%Furthermore, many \DRL algorithms assume that rewards are \emph{linear}, i.e.\ the algorithms attempt to maximize the sum of many small rewards within an episode. \change{Neither of these assumptions hold in the context of query latency: query latency is not dense, i.e. it can only be measured after the a plan has been formed, and query latency is not linear, i.e. just because many subtrees of an execution plan have a low cost does not mean that the entire execution plan has a low cost.}

One may reasonably consider using a traditional query optimizer's cost model as a reward signal instead of query latency, as the optimizer's cost model  may appear to provide a dense linear reward. {This approach has two major drawbacks. First,} these cost models tend to be complex, hand-tuned (by database engineers and DBAs) heuristics. Using  a cost model as the reward signal for a \DRL query optimizer simply ``kicks the can down the road,'' moving complexity and human effort {from designing optimization heuristics to tweaking optimizer cost models. Second,} the cost model's estimation of the quality of an execution plan may not always accurately represent the latency of the execution plan (e.g., a query with a high optimizer cost might outperform a query with lower optimizer cost). Therefore, using \DRL to find execution plans with a low cost as determined by a cost model might not always achieve the best possible results.

%Another candidate for a reward signal is the latency of the query itself: after an execution plan is selected, execute it, measure its latency, and use that value as a reward. This reward signal is not dense (it can only be computed after the terminal state), but it is linear and directly what we wish to minimize. When the reward signal can only be determined at the end of an episode, the reward signal is called ``sparse'' or ``delayed.'' Handling sparse reward signals is an active area of research in the machine learning community, with many algorithms (such as~\cite{ppo,trpo, dqn,ddqn,a3c}) capable of handling them. Nevertheless, delayed reward signals remain a challenge that researchers must keep in mind when selecting reinforcement learning schemes.

\sparagraph{Performance Evaluation Overhead} An often-unstated assumption made by many \DRL algorithms is that the reward of an action can be determined in constant time -- e.g., that determining the performance of an agent for a particular episode in which the agent performs poorly is no more time-consuming than calculating the reward  for an episode in which the agent performs well. For example, the time  to determine the current score of a player in a video game does not change based on whether or not the score is high or low.
If the latency of an execution plan is used as a reward signal, this assumption does not hold: poor execution plans can take \emph{significantly} longer to evaluate than good execution plans (hours vs.~seconds). Since traditional \DRL algorithms start with no information, their initial policies cannot be better than random choice, which will often result in very poor plans~\cite{howgood}. Hence, a naive \DRL approach that simply uses query latency as the reward signal would take a prohibitive amount of time to converge to good results.\footnote{\small We confirmed this experimentally by using query latency as the reward signal in ReJOIN. The initial query plans produced could not be executed in any reasonable amount of time.}

%Most query optimizer cost models have execution times that are relatively independent of the quality of the input execution plan. However, advanced query optimizers that utilize sampling or preemptive execution may have varying levels of independence between the quality of an execution plan and the cost model's runtime. 
 
%%% Local Variables:
%%% mode: latex
%%% TeX-master: "dlopt"
%%% End:
\nextSec
\section{Research Directions}
\label{sec:solutions}
Here, we outline potential approaches to handle the challenges we highlighted. First, we discuss two drastically different approaches, \emph{demonstration learning} and \emph{cost-model bootstrapping}, which both avoid the {pitfalls identified in Section~\ref{sec:challenges} in interesting ways.} We then touch upon \emph{incremental learning}, and propose three techniques that decompose the problem of query optimization in a principled way across various axes, and analyze the resulting design space.

%The common intuition behind all of these approaches is to split the learning process into phases, where early phases involve less complexity than later phases. By carefully incrementally increasing the complexity of the task in specific ways, the search space can be narrowed and the difficulties of the optimization metric/reward signal can be more easily managed.

%\subsection{Overcoming a troublesome optimization metric}
%\change{As described in Section~\ref{sec:reward_signal}, state-of-the-art \DRL algorithms depend on an objective metric/reward signal that is (1) measurable without a high cost and (2) incrementally dulled out over the course of an episode. Since query latency meets neither of these requirements (executing a bad query plan takes significantly longer than executing a good one, and query latency is only observable after a complete execution plan is built), this section describes two potential research directions that could help bridge this gap. The first direction, learning from demonstration, proposes uses techniques from the field of \DRL that initially learn by observing an expert~\cite{dqfd, demonstration}. The second direction, heuristic reward functions, describes how a traditional query optimizer's cost model might be used to ``bootstrap'' a more effective model.}

\subsection{Learning From Demonstration}

\begin{figure}
\centering
\includegraphics[width=0.3\textwidth]{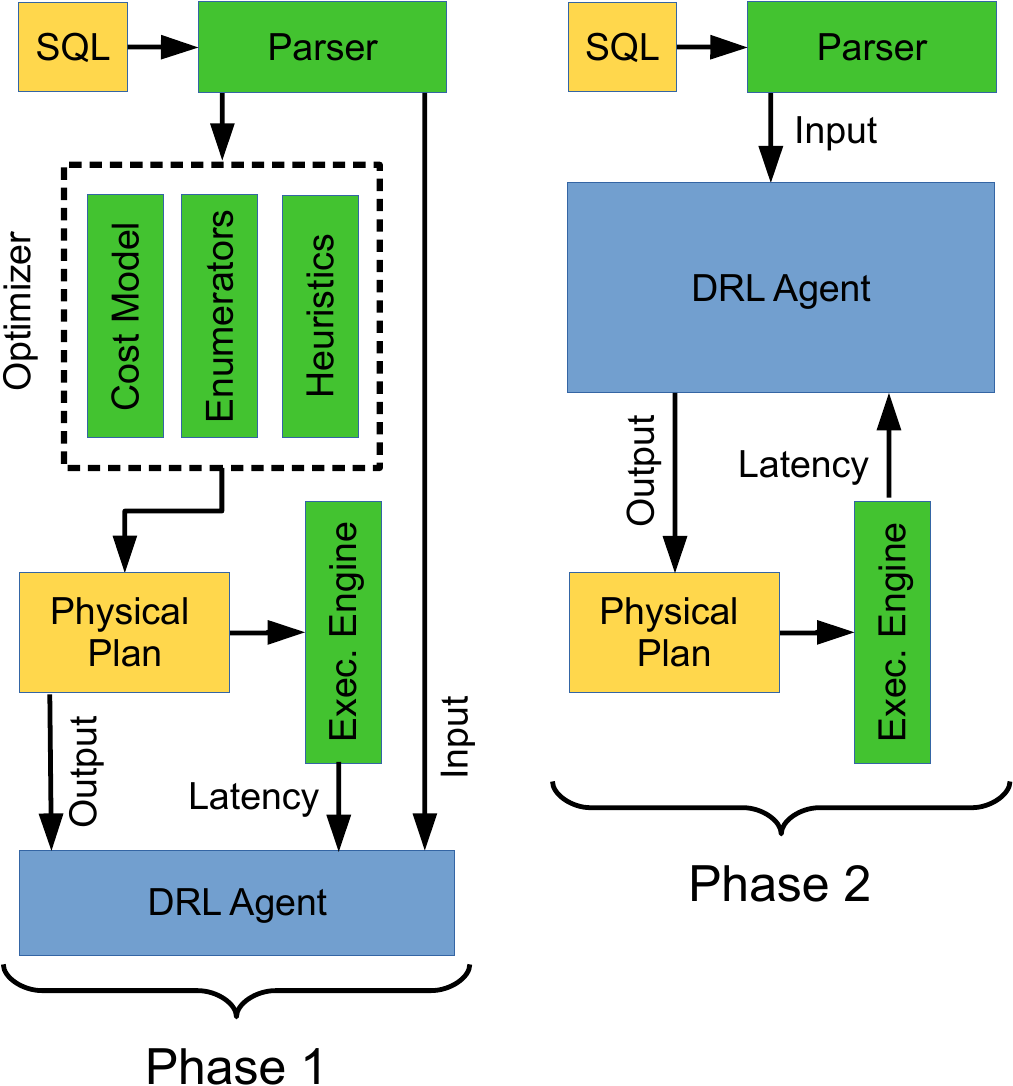}
\caption{\small Learning from demonstration}
\label{fig:demonstration}
\end{figure}

One way to avoid the pitfalls of using query latency directly as the performance indicator (reward) for \DRL algorithms is \emph{learning from demonstration} (LfD)~\cite{demonstration, dqfd}. Intuitively, this approach works by first training a model to imitate the behavior of an expert. Once this mimicry reaches acceptable levels, the model is fine-tuned by applying it to the actual environment. This learn-by-imitation technique mirrors how children learn basic behaviors like language and walking by watching adults, and then fine-tune those behaviors by practicing themselves.

Here, we propose using a traditional DBMS' query optimizer -- such as the PostgreSQL query optimizer -- as an expert. In this approach, illustrated in Figure~\ref{fig:demonstration}, a model is initially allowed to observe how the traditional query optimizer (the expert) optimizes a query. During this phase, the model is trained to mimic the optimizer's selected actions (e.g., indexes, join orderings, pruning of bad plans, etc). Assuming that a traditional optimizer will be able to prune-out unfeasible plans, this process allows a \DRL model to learn by observing the execution time of only feasible plans.

Once the model achieves good mimicry, it is then used to optimize queries directly, bypassing the optimizer. In this second phase, the model initially closely matches the actions of the traditional query optimizer, but now begins to slowly fine-tune itself based on the observed query latency. Here, the agent updates its neural network {based on the latency of the execution plans it constructs. If the performance of the model begins to slip, it is re-trained to match the traditional query optimizer until performance improves.} 
In practice, choosing the {point at which the model is again trained to mimic the traditional query optimizer} is critical to improve the performance of the algorithm~\cite{dqfd}. 
By leveraging learning from demonstration, one can train a query optimization model  that learns with small overhead, without having to execute a large number of bad plans, therefore  massively accelerating learning.  

While specific techniques and formalizations vary~\cite{dqfd, pretrain_demonstration, q, demonstration}, we outline the general process here.

\begin{enumerate}
\item{A large query workload, $W$, is executed one query at a time. Each $q \in W$ is transformed by the traditional query optimizer into a physical plan through a number of actions $a_i$ at various intermediary states $s_i$, which are recorded as an \emph{episode history}:
    $$H_q = [(a_0, s_0), (a_1, s_1), \dots, (a_n, s_n)]$$
    For example, at the initial state $s_0$, a query optimizer performing a greedy bottom-up join order selection process may choose an action $a_0$ signifying that two particular relations should be joined, or a query optimizer that first performs storage selection may choose an action signifying that data for a certain relation should come from a particular index. All episode histories are saved.}
\item{The resulting physical plans are executed, and the latency of each query $q \in W$, $L_q$, is measured and saved.}
\item{Next, the agent is trained, for each $q \in W$, on the $H_q$ and $L_q$ data (Phase 1 in Figure~\ref{fig:demonstration}). Specifically, for each action/state pair $(a_i, s_i) \in H_q$, the agent is taught to predict that taking action $a_i$ in state $s_i$ eventually results in a query latency of $L_q$.  Similar to the off-policy learning approach of~\cite{qo_state_rep}, the agent thus learns a \emph{reward prediction function}: a function that guesses the quality of a given action at a given state.}
\item{Once the agent has proficiency guessing the outcome of the traditional optimizer's actions, the agent can fine-tune itself. Now, the agent will be creating a query plan for an incoming query $q$. For a given state $s_i$, an action $a_i$ is selected by running every possible action though the reward prediction function and selecting the action which is predicated to result in the lowest latency.\footnote{\small In many implementations, an action besides the one predicted to result in the lowest latency may be selected with small probability~\cite{dqn} to enable additional exploration.} This process repeats until a physical execution plan is created and executed. The model is then trained (fine-tuned) on the resulting history $H_q$ and observed latency $L_q$.}
\item{Hopefully, the performance of the model will eventually exceed the performance of the traditional query optimizer. However, if the model's performance slips, it is partially re-trained with samples from the traditional query optimizer's choices when processing the queries in the initial workload $W$.}
\end{enumerate}

Since the behavior of the model in the second phase should not initially stray too far from the behavior of the expert system~\cite{dqfd}, we do not have to worry about executing any exceptionally poor query plans. Additionally, since the second training phase only needs to fine-tune an already-performant model, the delayed reward signal is of far less consequence. In fact, the initial behavior of the model may outpeform the traditional query optimizer in certain circumstances, for example if the trained model were to observe a systemic error in the performance of traditional optimizer, such as the traditional optimizer handling two similar situations in two significantly different ways, one of which causes substantially increased query latency. In this case, the trained model may automatically avoid the errors of the traditional optimizer (which has no capability to learn from its mistakes) \emph{through observation alone.}

An important issue here is that, since the experience collected based on the traditional optimizer is necessarily covering a narrow part of the action space (it excludes ``bad'' plans, and thus also excludes the corresponding sequence of actions that would produce them), many state-actions have never been observed and have no training data to ground them to realistic cost. For instance, a nested-loop-join or a table scan may never/rarely be picked by the traditional optimizer for a particular workload/database, and hence the model does not learn how to evaluate these actions correctly. {However, since the model is trained on experiences containing significantly faster execution plans, there is no reason for the model to attempt to explore these extremely poor plans.}

Experimental results from other problem domains (e.g.~arcade games~\cite{dqfd} and a few systems applications~\cite{lift}), show that deep reinforcement learning agents which initially learn from demonstration can master tasks with significantly less training time than their \emph{tabula rasa} counterparts. This result holds even when the expert is flawed (e.g.~when the expert is a human player who does not know a particular shortcut or strategy), implying that learning-from-demonstration techniques can improve upon, and not just imitate, existing expert systems.

\subsection{Cost Model Bootstrapping}
\begin{figure}
\centering
\includegraphics[width=0.43\textwidth]{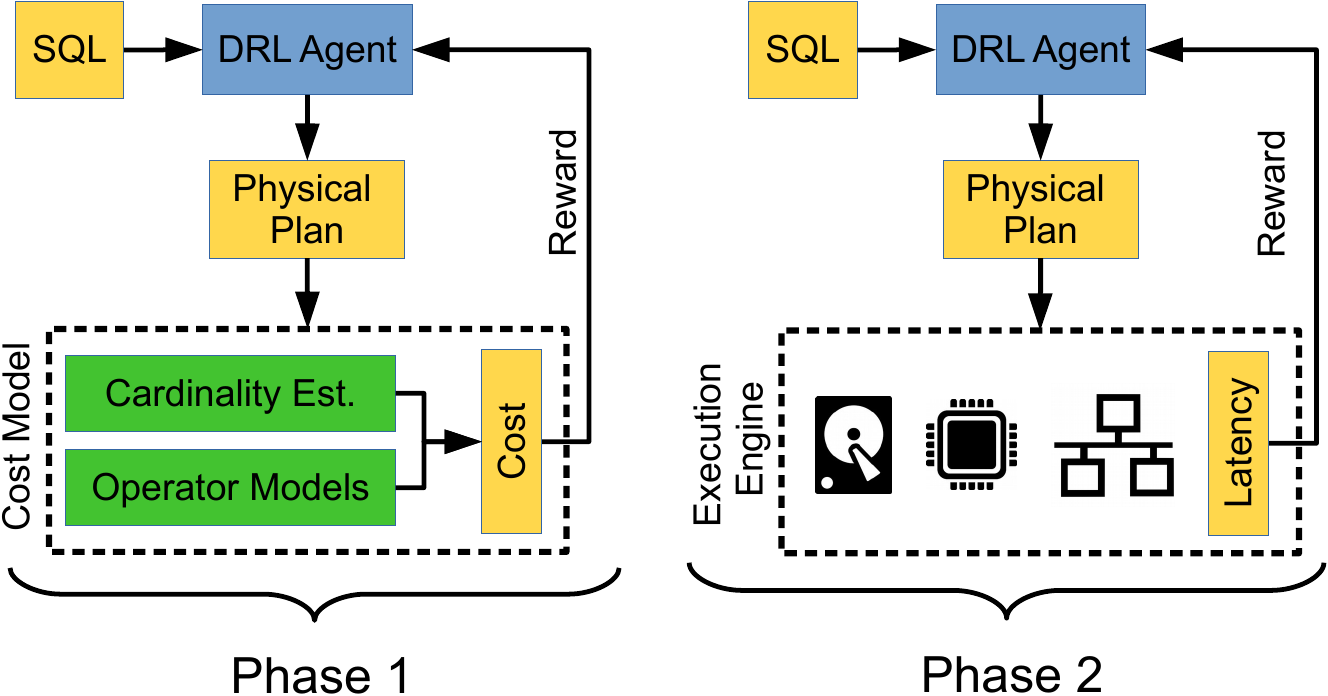}
\caption{\small Cost Model Bootstrapping}
\label{fig:bootstrap}
\vspace{-5mm}
\end{figure}

%One of the primary difficulties encountered by an end-to-end learning-based approach to query optimization is a cost function. Many reinforcement learning algorithms perform better when rewards are dulled out incrementally throughout an episode. However, in query optimization, the ``reward'' -- the latency of the resulting query plan -- is only known at the end of an episode.
A traditional, but still widely used and researched, approach to improving the performance of reinforcement learning algorithms on problems when the performance indicator (reward) is only available at the end of an episode (sparse) is to craft a \emph{heuristic reward function}. This heuristic reward function estimates the utility of a given state using a heuristic constructed by a human being: for example, when a robot is learning to navigate a maze, it may use an ``as-the-crow-flies'' heuristic to estimate its proximity to the maze's exit. In the game of chess, a popular heuristic to evaluate the value of a particular board position is to count the number of pieces captured by both sides.
Sometimes, this heuristic may be incorrect (e.g., it may rate a dead-end very near the exit as a desirable position, or it may highly-rate a board position in which many pieces have been captured but the opponent has an obvious winning move), but in general there is a strong relationship between the value of the heuristic function and the actual reward.

Luckily, the database community has invested significantly into designing optimizer cost models, which can be used for exactly this purpose. While imperfect, modern cost models, like ``as-the-crow-flies'' distance, can normally differentiate between good and catastrophic plans. We thus propose using these cost models as heuristic reward functions. This  approach, depicted in Figure~\ref{fig:bootstrap}, first uses the optimizer's cost model as a reward signal (Phase 1) and then, once training has converged, switches the reward signal to the observed query latency (Phase 2). In this way, the optimizer's cost model acts as ``training wheels,'' allowing the \DRL model to explore strategies that produce catastrophic query plans without requiring execution. Once the \DRL model has stabilized and starts to pick predominately good plans, the ``training wheels'' can be removed and the \DRL model can fine-tune itself using the ``true'' reward signal, query latency.

Cost model bootstrapping brings about a number of complications which require further exploration by the database community. Generally, an optimizer's cost model output is a unitless value, meant to compare alternative query plans but not meant to directly correlate with execution latency. For example, an optimizer's cost estimate for a set of query plans may range from 10 to 50, but the latency of these query plans may range from 100s to 200s. Switching the range of the reward signal from 10-50 to 100-200 will cause the \DRL model to assume that its performance has suddenly decreased (the \DRL model was getting query plans with costs in the range 10-50 in Phase 1, and at the start of Phase 2 the costs suddenly jump to be in range 100-200). This sudden change could cause the \DRL model to begin exploring previously-discarded strategies, requiring the execution of poor execution plans.  The change in variance could also have a detrimental effect~\cite{batchnorm}.

One way to potentially fix this issue would be to tune the units of the cost model to more precisely match execution latency, but the presence of cardinality estimation errors makes this difficult~\cite{howgood}. Instead of adjusting the optimizer's estimates to match the query latency, another approach could be to adjust the query latency to match the optimizer cost.  This could be implemented by simply scaling the query latency observed in Phase 2 to fall within the range of cost model estimates observed in Phase 1.

One could implement this scaling by noting the optimizer cost estimates \emph{and} query execution latencies during the end of Phase 1 (when the DRL model has converged). Let $C_{max}$ and $C_{min}$ be the maximum and minimum observed optimizer cost, and let $L_{max}$ and $L_{min}$ be the maximum and minimum observed query execution times. Then, in Phase 2, when the DRL model proposes an execution plan with an observed latency of $l$, the reward $r_l$ could be: 

\begin{equation*}
r_l = C_{min} + \frac{l - L_{min}}{L_{max} - L_{min}} (C_{max} - C_{min}) 
\end{equation*}

This scaling could be done linearly, as above, or using a more complex (but probably monotonic) function. This simple solution would likely need to be adjusted to handle workload shifts, changes in hardware, changes in physical design, etc.

Another potential approach, partially suggested in~\cite{sanjay_wat}, is to first train a neural network model to optimize for the operator cost, and then transfer the weights of the later layers of the network into a new network that trains directly on query latency. This technique, known as ``transfer learning'', has seen wide success in other fields~\cite{transfer, transfer2}.

%%% Local Variables:
%%% mode: latex
%%% TeX-master: "dlopt"
%%% End:

\nextSec
\subsection{Incremental Learning}
\label{sec:incremental}
In this section, we discuss potential techniques to \emph{incrementally} learn query optimization by first training a model to handle simple cases and slowly introducing more complexity. This approach makes the extremely large search space more manageable by dividing it into smaller pieces. Similar incremental approaches has shown success in other applications of reinforcement learning~\cite{dex, incremental_rl, transfer_rl}.

%Intuitively, incremental learning mirrors the process of human learning: one first achieves mastery in arithmetic, then algebra, and then calculus. 

%We then provide some insights into the design space of these potential decompositions. Using these insights, we propose and analyze three potential incremental learning approaches for end-to-end query optimization. 

\begin{figure}
\centering
\includegraphics[width=0.18\textwidth]{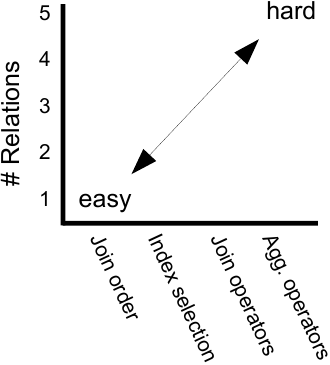}
\caption{\small Complexity diagram}
\label{fig:complexity_diagram}
\vspace{-5mm}
\end{figure}

We begin by examining how the task of query optimization can be decomposed into simpler pieces in a number of ways. 
We  note that the difficulty of a query optimization task is primarily controlled by two dimensions: the number of relations in the query, and the number of optimization tasks that need to be performed. This is illustrated in Figure~\ref{fig:complexity_diagram}. The first axis is the number of relations in the query.  If a \DRL model must optimize queries containing only a single relation, then the search space of query plans is very small (there are no join orderings or join operators to consider).  However, if a \DRL model must optimize queries containing many relations, then the search space is much larger. 

The second axis is the number of optimization tasks to perform. Consider a simplified query optimization pipeline (illustrated in Figure~\ref{fig:incremental}) containing four phases: join ordering, index selection, join operator selection, and aggregate operator selection. Performing any prefix of the pipeline is a simpler task than performing the entire pipeline: e.g., determining a join ordering and selecting indexes is a simpler task than determining a join ordering, selecting indexes, and determining join operators.

\begin{figure}
\centering
\includegraphics[width=0.5\textwidth]{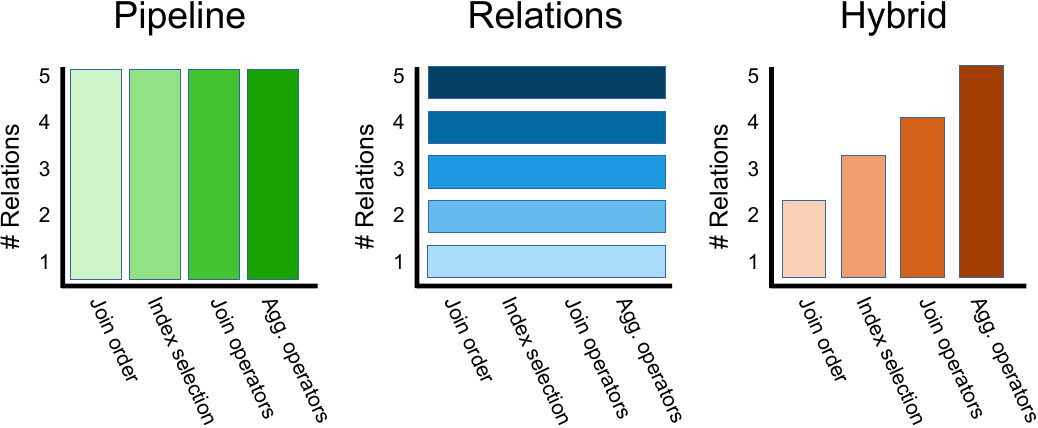}
\caption{\small Potential decompositions}
\label{fig:complexity_decompose}
\end{figure}

Thus, the lower-left hand side of Figure~\ref{fig:complexity_diagram} corresponds to ``easy'' cases, e.g. few stages of the pipeline and few relations. The upper-right hand side of Figure~\ref{fig:complexity_diagram} corresponds to ``hard'' cases, e.g. most stages of the pipeline and many relations.
This insight illuminates a large design space for incremental learning approaches. In general, an incremental learning approach will be divided into phases. The first phase will use ``easier'' cases (the bottom left-hand part of the chart), training until relatively good performance is achieved. Then, subsequent phases will introduce more complex examples to the model, allowing the model to slowly and smoothly learn more complex cases (the top right-hand part of the chart).

Figure~\ref{fig:complexity_decompose} illustrates three simple incremental learning approaches, {with light colors representating the initial training phases and dark colors representing the subsequent training phases.}
%(from left to right):
%\begin{itemize}
%\item{A \emph{pipeline stage}-based approach, in which the initial training phase (light green) requires only join ordering to be performed, and each subsequent  phase adds an additional component of the query optimization pipeline (darker greens).}
%\item{A \emph{relations}-based approach, in which the initial training phase involves only queries with few relations (light blue), and each subsequent phase permits a greater number of relations (darker blues).}
%\item{A \emph{hybrid} approach, which combines the previous two. The initial training phase uses only queries with few relations \emph{and} requires only join ordering to be performed (light red). Subsequent training phases both permit queries with a greater number of relations and adds an additional component to the query optimization pipeline (darker reds).}
%\end{itemize}
We next discuss each of these approaches in detail.

\subsubsection{Increasing optimization actions (pipeline)}
\begin{figure}
\centering
\includegraphics[width=0.38\textwidth]{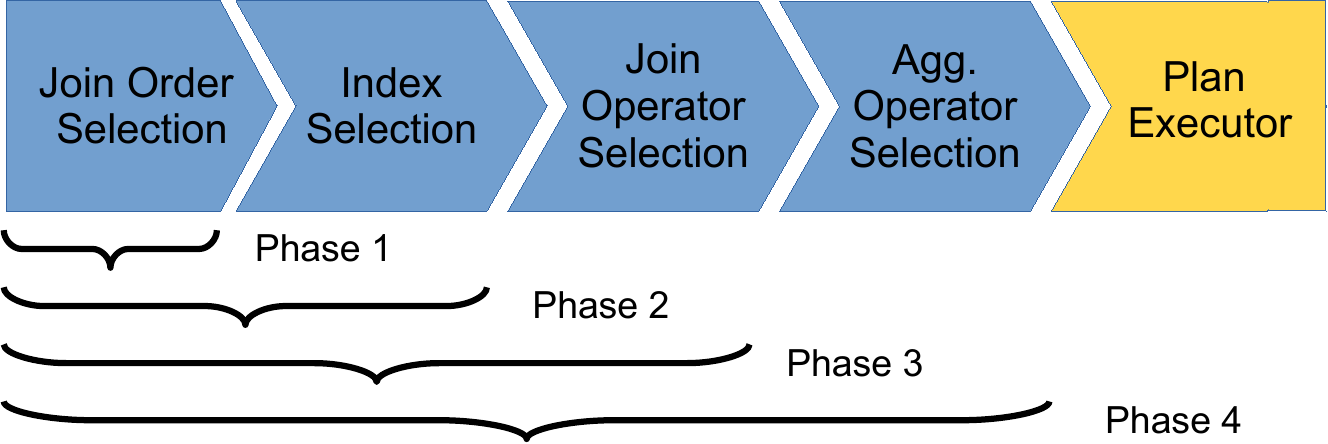}
\caption{\small Learning incrementally}
\label{fig:incremental}
\end{figure}

Our first proposed approach is \emph{pipeline-based incremental learning}, illustrated in Figure~\ref{fig:incremental}. A model is first trained on a small piece of the query optimization pipeline, e.g. join order selection. During this first phase, traditional query optimization techniques are used to take the output of the model and construct a complete execution plan (ReJOIN~\cite{rejoin} is essentially this first phase). Once the model achieves good performance in this first phase, the model is then slightly modified and trained on the first two phases of the query optimization pipeline, e.g. join order selection and index selection. This process is repeated until the model has learned the entire pipeline.

Extending ReJOIN to support this approach would be relatively straightforward. As shown in~\cite{rejoin}, the first phase of query optimization (join order enumeration) can be effectively learned. Once this initial training is complete, the action space can be extended to support index selection: instead of having one action per relation, the extended action space would have one action per relational data structure, e.g.~one action for a relation's B-tree index, one action for a relation's row-order storage, one action for a relation's hash index, etc. The knowledge gained from the previous training phase should help the model train significantly faster in subsequent phases.

The pipeline approach has the advantages of incremental learning (e.g., a managable growth of the state space), but comes with several drawbacks that need to be further investigated. First, the early training phases requires access to a traditional implementation of the later stages of the query optimization pipeline. While such implementations are available in a range of DBMSes today, the dependency on a traditional query optimizer is not ideal. Second, each phase of the training process will not bring about a uniform increase in complexity. It is conceivable that some stages of the pipeline are fundamentally more complex than others (for example, join order selection is likely more difficult than aggregate operator selection). The non-linearity of complexity going through the query optimization pipeline means that some training phases will require overcoming much larger jumps in complexity than others. This could result in unpredictable training times, or, in the worst case, a jump in complexity to large to learn all at once.

%\todo{w.r.t. the next paragraph commented out, do we intend to just not talk about the reward function for the incremental learning and assume the reader will think of cost model or cost model bootstrapping?}
%Identifying a good reward signal for each phase of the incremental training process is problematic. One potential but imperfect solution would be to use the optimizer's pre-existing cost model as a reward signal for all but the last phase of training, and then switch the reward signal to the latency of the query for the last phase. While poor decisions in the last phase could still result in poor query performance, even worst-case choices in the last phase of a query optimization pipeline (e.g. aggregate operator selection) have a much smaller impact on query performance than elements earlier in the pipeline (e.g. join order selection).
\subsubsection{Increasing relations}

\begin{figure}
\centering
\includegraphics[width=0.38\textwidth]{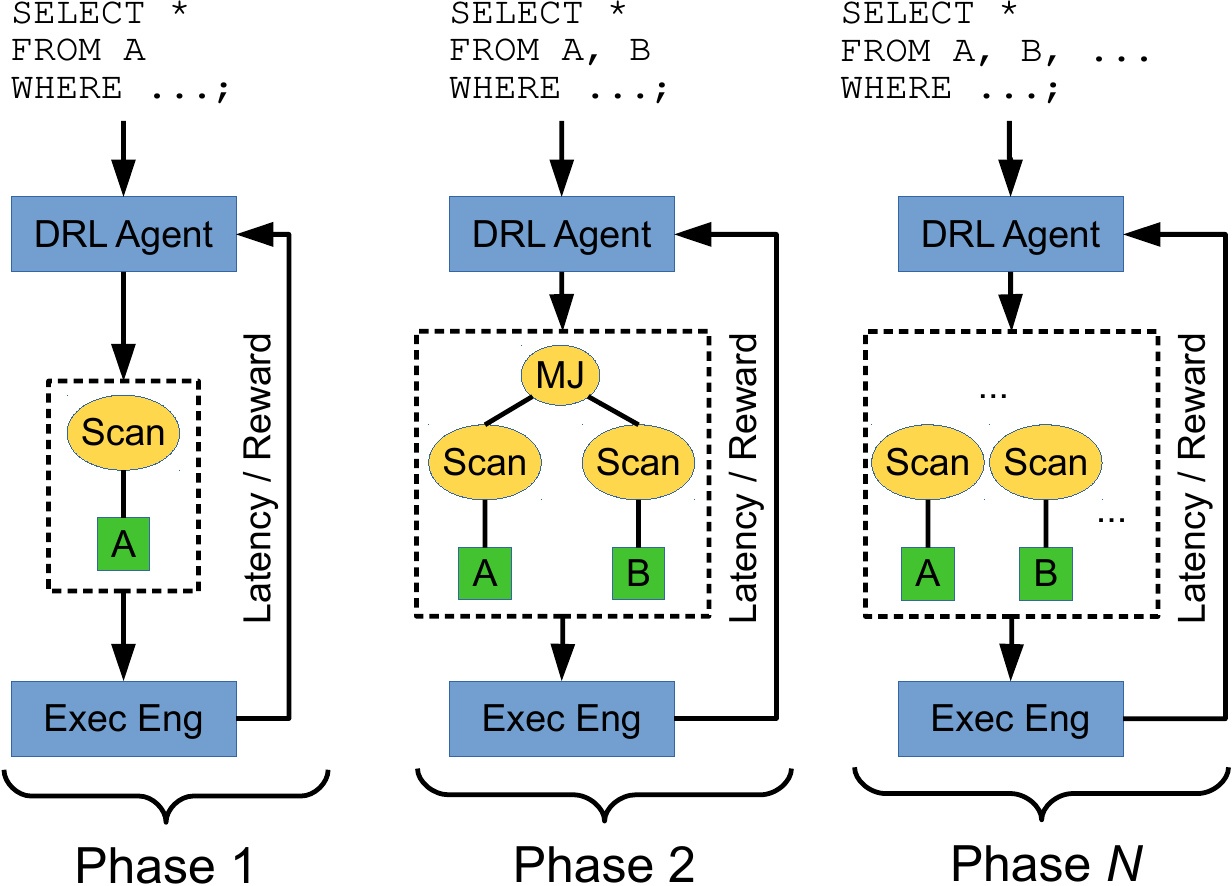}
\vspace{-2mm}
\caption{\small Learning from small examples}
\label{fig:small_examples}
\vspace{-5mm}
\end{figure}

While the previous approach reduces the size of the search space by focusing on larger and larger parts of the query optimization pipeline, this section proposes limiting the search space by focusing on larger and larger \emph{queries}. The proposed approach is depicted in  Figure~\ref{fig:small_examples}. In the first training phase, the model learns to master queries over a single relation. In subsequent training phases, the model is trained on queries over two relations, then three relations, etc. In each phase, the entire query optimization pipeline is performed.

This approach dodges some pitfalls of the pipeline stage approach. Generally, the increase in complexity between optimizing a query with $n$ relations and optimizing a query with $n+1$ relations is small. Even though there is an exponential increase in the number of potential join orderings, this is a ``quantitative'' change as opposed to a ``qualitative'' change -- intuitively, it is easier to learn how to create a join plan with a single additional relation than it is to learn how to perform a new pipeline step.

A major challenge of this approach is finding candidate queries. Generally, real-world workloads will contain very few queries over a single relation. Even synthetic workloads have very few low-relation-count queries (TPC-H~\cite{tpch} has only two such templates, JOB~\cite{howgood} has none). Queries with low relation counts could be synthetically generated, but doing so while matching the characteristics of real-world workloads is a complex task.

\subsubsection{Hybrid}
The last approach we explicitly discuss is the hybrid approach, depicted on the right-hand side of Figure~\ref{fig:complexity_decompose}. In this hybrid approach, the initial training phase learns only the first step of the query optimization pipeline (e.g. join order selection) using only queries over two or fewer relations. The next training phase introduces both another step of the pipeline (e.g. index selection) \emph{and} queries over three or fewer relations. After all stages of the query optimization pipeline have been incorporated, subsequent training phases increase the number of relations considered. This approach provides the smallest increase in complexity from training phase to subsequent training phase. However, the hybrid approach suffers from some of the disadvantages of both the relations and pipeline based approach: it depends on a traditional optimizer and it requires queries with relatively few relations for training purposes.

%One can easily imagine many more ways to incrementally increase complexity than are depicted in Figure~\ref{fig:complexity_decompose}. For example, one could increase the number of relations considered multiple times before considering another pipeline step. One could increase the number of relations considered multiple times, and then decrease the number of relations considered when adding a new pipeline step. Each of these strategies come with their own advantages and disadvantages, leading to many possible research directions.

%%% Local Variables:
%%% mode: latex
%%% TeX-master: "dlopt"
%%% End:

%%% Local Variables:
%%% mode: latex
%%% TeX-master: "dlopt"
%%% End:
\nextSec
\section{Conclusions}
\label{sec:conclusions}

We have argued that recent advances in deep reinforcement learning open up new research avenues towards a ``hands-free'' query optimizer, potentially improving the speed of relational queries and significantly reducing time spent tuning heuristics by both DBMS designers and DBAs. We have identified how the large search space, delayed reward signal, and costly performance indicators provide substantial hurdles to naive applications of \DRL to query optimization. Finally, we have analyzed how recent advances in reinforcement learning, from learning from demonstration to bootstrapping to incremental learning, open up new research directions for directly addressing these challenges.

\sparagraph{Other complexities} We argue that deep reinforcement learning can greatly decrease the amount of human effort required to develop and tune database management systems. However, these deep learning techniques come with their own complexities as well: training configurations (e.g. learning rate), network architectures, activation function selection, etc. While deep learning researchers are quickly making inroads towards automating many of these decisions~\cite{rl_arch, evo_nn}, future research should carefully analyze the tradeoffs between tuning deep learning systems and tuning traditional query optimizers. 

\sparagraph{Other applications} While query optimization is a good candidate for applying \DRL to database internals, a wide variety of other core DBMS concepts (e.g.cache management, concurrency control) could benefit from applications of machine learning as well. Careful applications of machine learning across the entire DBMS, not just the query optimizer, could bring about a massive increase in performance and capability. 
%%% Local Variables:
%%% mode: latex
%%% TeX-master: "dlopt"
%%% End:

\small
\bibliographystyle{acm}
\bibliography{ryan-cites-short}

%\appendix

\end{document}